\documentclass[runningheads]{llncs}
\usepackage{cite}
\usepackage{amsmath,amssymb,amsfonts}
\usepackage{algorithmic}
\usepackage{graphicx}
\usepackage{textcomp}
\usepackage[letterpaper]{geometry}
\usepackage{times}
\usepackage[none]{hyphenat}
\usepackage{url}
\usepackage{latexsym}
\usepackage{indentfirst}
\usepackage{graphicx}
\usepackage{epsfig}
\usepackage{subcaption}
\usepackage{calc}
\usepackage{amssymb}
\usepackage{amstext}
\usepackage{amsmath}
\usepackage{amsthm}
\usepackage{multicol}
\usepackage{pslatex}
\usepackage{diagbox}
\usepackage{cite}
\usepackage[dvipsnames]{xcolor}
\usepackage{multirow}
\usepackage{url}
\usepackage{subcaption}
\usepackage{hyperref}
\usepackage[T1]{fontenc}
\usepackage[ruled,vlined]{algorithm2e}
\emergencystretch=\maxdimen
\begin{document}
\title{A framework for syntactic and semantic quality evaluation of ontologies}

\author{Vivek Iyer\inst{1,2}\orcidID{0000-0003-4451-8293} \and
Lalit Mohan Sanagavarapu\inst{1,2}\orcidID{0000-0003-0745-1042} \and
Y Raghu Reddy\inst{1,3}\orcidID{0000-0003-2280-5400}}
\authorrunning{V. Iyer et al.}
\institute{IIIT Hyderabad, India\and
\email{\{vivek.iyer,lalit.mohan\}@research.iiit.ac.in}\and
\email{raghu.reddy@iiit.ac.in}}

\maketitle              % typeset the header of the contribution
\begin{abstract}
The increasing focus on Web 3.0 is leading to automated creation and enrichment of ontologies and other linked datasets. Alongside automation, quality evaluation of enriched ontologies can impact software reliability and reuse. Current quality evaluation approaches oftentimes seek to evaluate ontologies in either syntactic (degree of following ontology development guidelines) or semantic (degree of semantic validity of enriched concepts/relations) aspects. This paper proposes an ontology quality evaluation framework consisting of: (a) SynEvaluator and (b) SemValidator for evaluating syntactic and semantic aspects of ontologies respectively. SynEvaluator allows dynamic task-specific creation and updation of syntactic rules at run-time without any need for programming. SemValidator uses Twitter-based expertise of validators for semantic evaluation. The efficacy and validity of the framework is shown empirically on multiple ontologies.

\end{abstract}
\section{Introduction}
The exponential increase in Internet users over the past decade has led to generation of large volume of data. Web 3.0, otherwise commonly referred to as Semantic web, seeks to represent internet data as knowledge through knowledge graphs, ontologies and other knowledge systems \cite{berners2001semantic}. 
% These representations enable integration of mixed sources, dissolve ambiguities in word or phrases, identify relevant information, provide decision support, reasoning and many other use cases. 
These representations enable knowledge integration, semantic ambiguity resolution, information extraction, decision making, reasoning and many other use cases relevant to the building of `intelligent' software systems. Ontologies, in particular, store domain-specific knowledge, and represent this knowledge through concepts, relations, axioms and instances. They contain a formal structure and achieve a certain level of rigor due to the presence of rules and constraints. %Knowledge graphs \cite{paulheim2017knowledge}, on the other hand, are %domain-agnostic and have gained popularity over the last decade for %representing data across various domains.
Ontologies are rarely static in nature. The range and the depth of the knowledge stored are enriched over time. This impacts a wide variety of software applications that utilize ontologies for reasoning, decision-making, question-answering, etc. Ontology enrichment is thus a crucial step in the ontology engineering process. 
%The previously mentioned explosion in internet data has also le`d to a %large body of work dedicated to enriching both ontologies and %knowledge graphs.
%

Traditionally, ontologies are created and managed by knowledge engineers and domain experts resulting in high costs due to the expert human labour involved. Automated or semi-automated approaches to ontology enrichment are increasingly popular, driven by increased availability of domain-relevant internet data and improvements in natural language processing and machine learning models \cite{wong2012ontology}. Research on ontology learning (both creation and enrichment) snowballed in the last two decades \cite{maedche2001ontology}, \cite{wong2012ontology}, \cite{iyer2019survey}, with increased focus on fully automated Deep Learning based approaches \cite{sanagavarapu2021ontoenricher}. Given the variety in ontology enrichment approaches, it is important to evaluate the quality of the enriched ontology.
%============
\begin{figure}[ht]
  \centering
  % include first image
  \includegraphics[width=0.7\textwidth]{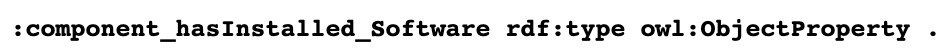}  
  \caption{Syntactic quality violation: No domain or range for property}
  \label{fig:syntactic-quality}
\end{figure}
%==============
%==============
\begin{figure}[ht]
  \centering
  % include first image
  \includegraphics[height=150px]{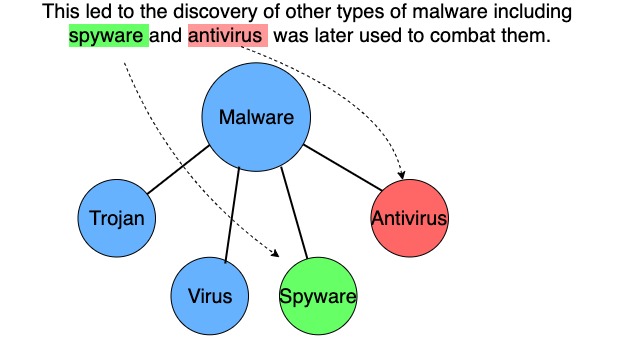}  
  \caption{Semantic violation: Invalid enriched concept}
  \label{fig:semantic-quality}
\end{figure}
%================
 
Ontology evaluation approaches can broadly be divided into: manual, automated and semi-automated approaches. In general, ontology evaluation happens on one of two aspects: syntactic quality, or semantic quality. We define syntactic quality of an ontology as a measure of its adherence to ontology development guidelines or rules. For example, One such rule could necessitate presence of both domain and range elements in properties. Examples of other rules or guidelines could include explicit declaration of equivalent and inverse properties, presence of annotations \cite{vrandevcic2009ontology}, following of unique naming conventions \cite{noy2001ontology}, etc. Figure \ref{fig:syntactic-quality} shows, in Turtle syntax \cite{beckett2014rdf}, a property without a defined range element - thus violating the rule that necessitates the presence of both domain and range elements. Semantic quality deals with validity of enriched concepts, relations and instances. Figure \ref{fig:semantic-quality} shows an example ontology enriched with concepts extracted from a sentence to emphasize the need for evaluation of enriched ontologies. The previously existing concepts are shown in blue, the valid enriched concepts in green and the invalid enriched ones in red. Ontology enrichment algorithms using Hearst patterns \cite{hearst1992automatic} could mistakenly detect `antivirus' as a type of `malware'.  In such cases, before creating a final ontology, the enriched ontology needs to be validated for semantic quality, to accept or reject the enriched concepts, properties and instances. 

A variety of metric-based methods were proposed to evaluate various syntactic quality-based characteristics of ontologies \cite{duque2013evaluation}, \cite{lozano2004ontometric}, \cite{tartir2010ontological}. 
% The works of Vrandevic \cite{vrandevcic2009ontology}, Astrid et al. \cite{duque2013evaluation}, Pammer et al. \cite{pammer2006ontology} and Noy et al. \cite{noy2013getting} suggested approaches and metrics that include `completeness' (software engineering quality principle) as a measure to syntactically evaluate ontologies. 
Publicly available tools were proposed that allow users to evaluate syntactical quality of ontologies using pre-defined metrics \cite{lantow2016ontometrics}, \cite{mcdaniel2018assessing} or rules \cite{poveda2012validating}, \cite{schober2012ontocheck}. However, they offer limited customization and flexibility to the user for creating task-specific rules for evaluation, even more so for non-programmers. In regards to semantic evaluation, researchers have traditionally employed domain experts \cite{makki2008nlp}, while in this decade, crowdsourced validators \cite{kontokostas2013triplecheckmate}, \cite{noy2013mechanical} are being used for semantic ontology validation. This paper proposes a customizable and scalable framework that evaluates syntactic and semantic aspects of ontology quality using SynEvaluator and SemValidator respectively.
\begin{itemize}
    \item $SynEvaluator$: a tool that uses a rule-creation framework for allowing users to non-programatically create rules during run-time and to set task-specific priorities for these rules.
    \item $SemValidator$: a tool that uses crowdsourcing for validation of semantic quality of enriched ontologies. In this paper, a Twitter-based expertise estimation algorithm is used to weight validators' decisions.
\end{itemize}  

The source code for SynEvaluator and SemValidator is available on GitHub\footnote{https://github.com/Remorax/SynEvaluator/}\textsuperscript{,}\footnote{https://github.com/Remorax/SemValidator/}. They are also deployed on Heroku as web-applications\footnote{https://synevaluator.herokuapp.com/}\textsuperscript{,}\footnote{http://semvalidator.herokuapp.com/}. 

The rest of the paper is structured as follows: Section 2 describes related work in ontology quality evaluation. The proposed framework constituting of SynEvaluator and SemValidator is shown in section 3. Section 4 details the experiments done to show the efficacy and accuracy of the framework, through these tools. SynEvaluator is tested for its utility and accuracy for implementing syntactical quality evaluation rules by comparing it against a popular syntactic quality evaluation tool, OOPS! \cite{poveda2018oops}. The efficacy of $SemValidator$ is shown by conducting crowdsourced survey involving 28 validators on two popular ontologies, Stanford Pizza  \cite{StanfordPizza} and Information Security ontology \cite{SteFenz}. Accuracy of TweetExpert algorithm on responses to both of these ontologies using multiple ML regression algorithms is shown. Finally, section 5 summarizes the contributions and suggests possible future directions of research.

\section{Related Work}
% Ontology evaluation can be conducted: (a) on a stand alone basis (b) with some context (c) within an application, and (d) in the context of an application and a task \cite{vrandevcic2009ontology}. Ontology evaluation efforts can also be divided into `black box' or `task based' strategies, and `glass box' or `component' strategies. Black box strategies are primarily used from an end-user perspective. Glass box strategies are based on individual characteristics of the ontology and are applied throughout its life cycle \cite{hartmann2005d1}. 

% Another classification, by Brank et al. \cite{brank2005survey} is based on two dimensions (i) type of approach (comparison against a gold standard, application or task-based evaluation, user based evaluation, and data-driven evaluation) and (ii) level of evaluation (lexical, vocabulary, or data layer; hierarchy, taxonomy; other semantic relations; context, application; structure, architecture, design). %
Ontology quality evaluation approaches can be broadly classified into a) syntactic and b) semantic quality evaluation approaches.
Syntactic evaluation approaches primarily evaluate structural aspects of an ontology based on ontology development guidelines, common pitfalls, structural metrics etc. OntoClean \cite{guarino2002evaluating}, one of the earliest known works in this area, proposed a methodology for validating adequacy of relationships in an ontology based on notions drawn from philosophy such as essence, identity, and unity. Similarly, OntoQA \cite{tartir2010ontological} proposed ontology evaluation on the basis of schema metrics and instance metrics. They stated that `goodness' or `validity' of an ontology varies between different users and domains.   Gangemi et al. \cite{gangemi2005theoretical} proposed structural, functional and usability-related measures using O$^2$ and oQual, a meta-ontology and an ontology model for ontology evaluation and validation. Burton et al. \cite{burton2005semiotic} proposed an ontology evaluation model based on semiotic theory. In order to apply the metrics proposed in these works, tools such as S-OntoEval \cite{dividino2008semiotic} drawn from semiotic theory and AktiveRank \cite{alani2006ranking} that ranked ontologies based on structural metrics like class match measure, density measure etc. were proposed. However, the tools proposed in these articles are either closed-source prototypes or theoretical frameworks and are not publicly available. 

There are also a few publicly available closed-source ontology (syntactic) evaluation tools, such as OOPS! \cite{poveda2012validating}, DoORS \cite{mcdaniel2018assessing} and OntoMetrics \cite{lantow2016ontometrics}. OntOlogy Pitfall Scanner (OOPS!) evaluates ontologies on the basis of established ontology quality rules related to human understanding, logical consistency, real world representation and modelling issues, and manually assigned priorities. DoORS, evaluates ontologies based on metrics drawn from Semiotic Theory while OntoMetrics uses metrics proposed in OntoQA, \cite{gangemi2005theoretical}. These tools, however are not flexible or customizable, and evaluate ontologies using a fixed set of rules or metrics. Users are unable to create/update customized rules and set task-specific priorities, which can be crucial as requirements and application scenarios vary. The available open source implementations \cite{ontoevalimplementation} require creation of new rules and priorities via programming which can be daunting for non-programmers. 

Semantic evaluation approaches focus on semantic validity of concepts and relationships in an ontology. Traditionally, it has been formulated as a task requiring simple accept/reject decisions from domain experts \cite{amardeilh2005document}. In the past few years, a good number of crowdsourced ontology evaluation approaches have been proposed. Hanika et al. \cite{wohlgenannt2016crowd} have developed the UComp Protégé plugin to provide a platform for crowdsourced workers to validate classes, subclasses, properties and instances. Kiptoo et al. \cite{kiptoo2020ontology} use crowdsourcing for axiom and assertion verification in ontologies as well as for verification of subclass-superclass relations by Amazon Mechanical Turks \cite{noy2013mechanical}. Pittet et al. used crowdsourced workers to propose changes related to addition, deletion and substitution errors \cite{pittet2015exploiting}. Zhang et al. \cite{zhang2017semantic} used crowdsourced workers to obtain written feedback (comments/suggestions/references) for making final validation decisions. Requiring complex tasks (such as making data quality decisions or requiring written feedback) from crowdsourced workers can be expensive and unscalable as the size of ontology and/or number of workers increases. Some approaches \cite{wohlgenannt2016crowd}, \cite{kontokostas2013triplecheckmate}, \cite{pittet2015exploiting} used quality control mechanisms like majority voting that are debatable. Noy et al. \cite{noy2013mechanical} addressed this by using qualification questions and spam filtering techniques. While these mechanisms eliminate spammers, it may not be applicable where large number of workers have some degree of knowledge but only few are experts. Therefore, an assessment of domain expertise on a continuous scale would be useful as a quality control metric. Further, an integrated quality evaluation framework that seeks to evaluate ontologies on both syntactic and semantic aspects, and also addresses the above-mentioned problems would be useful for a holistic and integrated approach to quality evaluation of enriched ontologies.
\section{Proposed Framework}
In this paper, an ontology evaluation framework that combines automated syntactic evaluation and human-centric semantic evaluation is proposed. SynEvaluator aims at increasing user flexibility by allowing customized rule creation at runtime, as well as scalability (with respect to user base) by proposing an approach that removes the need for programming. SemValidator proposes a crowdsourcing based approach that uses the validators' Twitter profiles for quality control. 

\begin{figure*}[h!]
    \centering
    {\includegraphics[width=\textwidth]{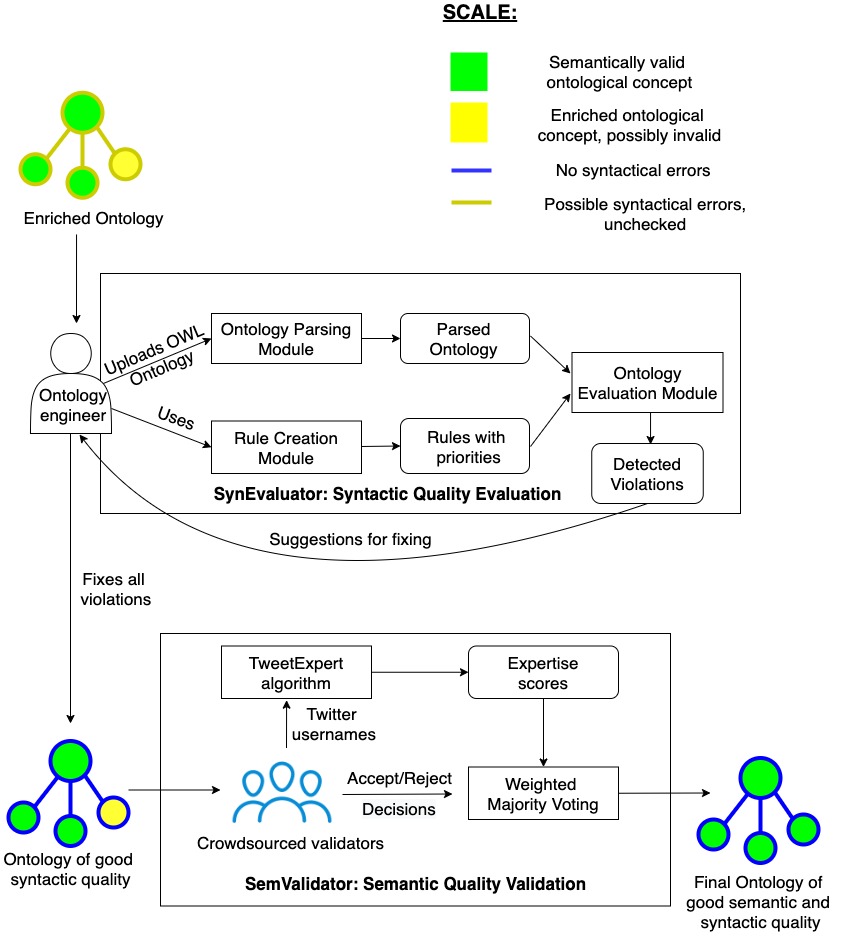}}
    \caption{Work flow for semantic and syntactic quality evaluation of ontologies.}
    \label{fig:workflowdiagram}
\end{figure*}

The framework's work flow is shown in Figure \ref{fig:workflowdiagram}. The input to quality evaluation process is an enriched ontology. The ontology may have been enriched with concepts, relations and instances using some automated and semi-automated algorithms. The ontology may contain syntactic quality errors (due to violation of ontology development guidelines) and/or semantic errors (due to wrongly enriched concepts). In the figure, for clarity, concepts and relationships with potential syntactic violations are bordered in yellow, an those that do not contain such violations are outlined in blue. 
% Note that both concepts and relations (as well as instances and other elements of the ontology) could contain syntactic violations such as missing domain and range, missing annotations, wrong inverse/equivalent relationships etc. 
Also, concepts that potentially contain semantic violations are highlighted in yellow colour and concepts that are semantically valid are highlighted in green.
% Please note that although any concept, relation or instance could contain a semantic quality violation, for simplicity's sake, we only highlight invalid concepts in the example. 

An ontology engineer uploads an enriched ontology to SynEvaluator, and creates syntactic quality evaluation rules. The rules created using a theoretical rule creation framework are applied on the parsed ontology object through SynEvaluator's ontology evaluation module. This returns a list of detected violations and the elements causing these violations. Using these elements as suggestions, the ontology engineer can fix violations in a iterative manner. The iterations may be repeated as needed. Then, the ontology engineer can provide this ontology as input to SemValidator so that it can be validated for semantic quality. As part of semantic validation, the ontology is provided to crowdsourced validators who give their accept or reject validation decisions for each of the enriched concepts, relations and instances. Simultaneously, an estimate of the domain knowledge of each of these validators is calculated from their Twitter profile using the TweetExpert algorithm. These scores alluded to as `TweetExpert scores', are used as a quality control mechanism to ensure that the decisions of crowdsourced validators are given weightages according to their knowledge of the ontology domain. Finally, the output of this algorithm is used to take the final accept/reject decision for each enriched element, resulting in an ontology with both good syntactic and semantic quality. 

\subsection{Stage 1: SynEvaluator}
In this section, the underlying terminology used in SynEvaluator and rule creation framework is illustrated with examples.  Further, the implementation of SynEvaluator as a web application is detailed. Finally, the section ends with a discussion on potential benefits and limitations of SynEvaluator. 

\subsubsection{Defining the Rule Creation Framework}

SynEvaluator allows users to create rules at run-time. The rules are constructed from individual components like ``Subjects", ``Clauses" or ``Operator Expressions". The operational definitions of these and other relevant terms are : 
\begin{itemize}
\setlength{\itemsep}{0pt}
\setlength{\parskip}{0pt}
\setlength{\parsep}{0pt}
    \item \textbf{Ontological Element}: Refers to any element that forms a constituent part of an ontology. It refers to any primary component (classes, individuals, properties), their related elements (subclasses, domains, ranges), annotations (labels, descriptions, comments) or attributes (ID, language, namespace).
    \item \textbf{Rule}: Refers to a sequence of one or more clauses, optionally connected by one or more operator expressions that returns either one or more ontological elements or a boolean value.
    \item \textbf{Clause}: Refers to a transformation applied on an ontological element(s) to return either one or more ontological elements or a boolean value.
    \item \textbf{Operator Expression}: An expression used to compare and/or connect non-empty sequences of clauses to return a boolean value.
    \item \textbf{Subject}: Refers to an ontological element (typically primary components such as classes, individuals and properties), that is subjected to transformations carried out through sequential clauses to form a rule.
\end{itemize}

\begin{figure}[h!]
    \centering
    {\includegraphics[width=0.5\textwidth]{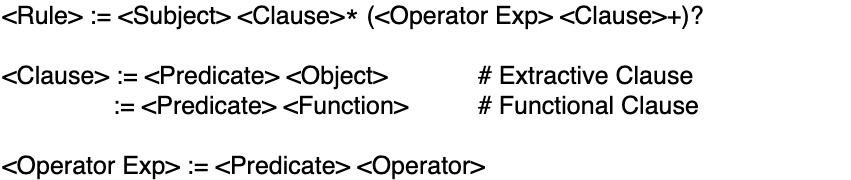}}
    \caption{Structure of supported expressions}
    \label{fig:definition-expression}
\end{figure}

Using these concepts, the expressions supported by SynEvaluator are formally defined as shown in Figure \ref{fig:definition-expression}. The notation used is similar to $RegEx$ notation with $*$ denoting zero to infinity, $+$ denoting one to infinity, and $?$ denoting zero or one occurrences. A `Subject' comprises the beginning and is always the first keyword in a rule in out proposed framework. It goes through a series of transformations as defined by sequences of clauses. Clauses can be of of two types: a) Extractive Clauses and b) Functional Clauses. Extractive Clauses consist of (Predicate, Object) pairs that use the Predicate to execute a transformation on the return value from the previous clause using the Object as argument. More specifically, object specifies the type of element `extracted'  by the predicate, and elements satisfying this (Predicate, Object) pair are returned as output. Functional clauses, on the other hand, consist of (Predicate, Function) pairs that involve executing a function of type described by predicate on the return value from the previous clause. These clauses typically check for existence of a certain functional property and thus return a boolean value in response. 

Currently, two kinds of functional properties are supported: i) ontological (or structural) properties, that execute ontology-level functions (such as uniqueness, validity, consistency etc) on ontological elements, and ii) linguistic properties that linguistically analyze text (such as checking for polysemes, conjunctions etc) returned from previous clauses. In case of `False' value returned by a Functional Clause or an empty list (no matching elements) returned by an Extractive Clause, that particular ontological element is returned as an element containing a violation.

Extractive Clauses can also be chained together to form clause sequences. Since functional clauses return a boolean value, they cannot be chained further. Clause sequences can be compared through Operator Expressions. Operator Expressions essentially consist of a `Predicate' indicating operator type, followed by an `Operator' keyword. Operator Expressions comprise of two main categories of operators, namely: (a) Logical Operators and (b) Comparative Operators. Logical Operators like `And', `Or' and `Not' are used to create logical combinations of clause sequences. Comparative Operators like `Equality', `Inverse' and `Synonymy' are used to compare return values.
\begin{table*}[h!]
\centering
\caption{Keywords for different expression types in the proposed framework}
\label{tab:keyword-examples}
\setlength\tabcolsep{3pt}
\begin{tabular}{|c|c|c|c|}
\hline
\multirow{2}{*}{\textbf{Subject}} & \textbf{Extractive Clause} & \textbf{Functional Clause} & \textbf{Operator Expression} \\ \cline{2-4} 
 & \textbf{Predicate} & \textbf{Predicate} & \textbf{Predicate} \\ \hline
Ontology Metadata & Has Related Element (1) & Has Ontological Property (1) & Uses Comparative Operator (1) \\ \hline
Ontological Element & Has Attribute (2) & Has Linguistic Property (2) & Uses Logical Operator (2) \\ \hline
Class & \textbf{Object} & \textbf{Function} & \textbf{Operator} \\ \hline
Instance & Domain (1) & ID Consistency (1) & Equality (1) \\ \hline
Property & Subclass (1) & Uniqueness (1) & Inverse (1) \\ \hline
Object Property & Disjoint Class (1) & Text Validity (1) & And (2) \\ \hline
Datatype Property & ID (2) & Contains Polysemes (2) & Or (2) \\ \hline
Symmetric Property & Language (2) & Contains Conjunctions (2) & Not (2) \\ \hline
\end{tabular}
\end{table*}

Table \ref{tab:keyword-examples} summarizes keywords supported by the proposed framework. Note that the table lists 8 Subject keywords, while for Extractive Clauses, Functional clauses and operator expressions it lists 2 predicates and 5 objects, functions and operators respectively. This is due to differing syntax followed by each expression type. Also, every predicate has a list of valid Objects/Functions/Operators. This is shown in the table through bracketed numbering. For example, the valid Predicates for `Has Attribute` are `ID` and `Language`. The complete list of supported keywords is provided over here\footnote{https://bit.ly/3zfdI8f}. 
\subsubsection{Examples of Created Rules}
\begin{figure*}[h!]
    \centering
    {\includegraphics[width=\textwidth]{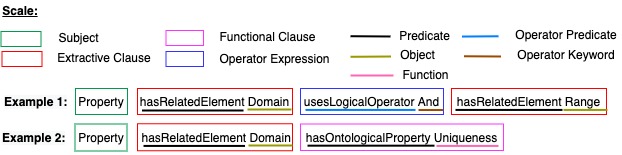}}
    \caption{Examples of rules implemented by SynEvaluator}
    \label{fig:examples-rules}
\end{figure*}

Figure \ref{fig:examples-rules} shows examples of 2 rules. In both examples 1 and 2, `Property' is the `Subject' of the rule. `hasRelatedElement Domain', `hasRelatedElement Range' and `hasOntologicalProperty Uniqueness' are all clauses. In rule 1, the `hasRelatedElement Domain' clause carries out a transformation that uses the `hasRelatedElement' predicate to extract `Domain' elements. A similar clause is used for extracting `Range' elements. These two clauses (or clause sequences of length are combined using the operator expression `usesLogicalOperator And'. This rule thus necessitates non-null values for both Domain and Range elements for each element of `Subject', in this case, `Property'. Properties that do not contain both elements are therefore returned as ontological elements containing violations.

Example 2 shows a rule where clauses have been chained together to constitute a clause sequence of length 2. This sequence consists of the extractive clause `hasRelatedElement Domain' followed by functional clause `hasOntologicalProperty Uniqueness'. The extractive clause essentially extracts Domain element(s) of each Property. Then, the functional clause applies `Uniqueness' function on ontological elements returned by previous clause with function type defined by Predicate ``hasOntologicalProperty". In case of non-existence of domain for a particular property or existence of multiple domains, this rule would return that property as containing a violation due to violation in first and second clauses respectively. 
\subsubsection{Proposing SynEvaluator: the final web application}

 \begin{figure}[h]
    \centering
    {\includegraphics[width=\linewidth]{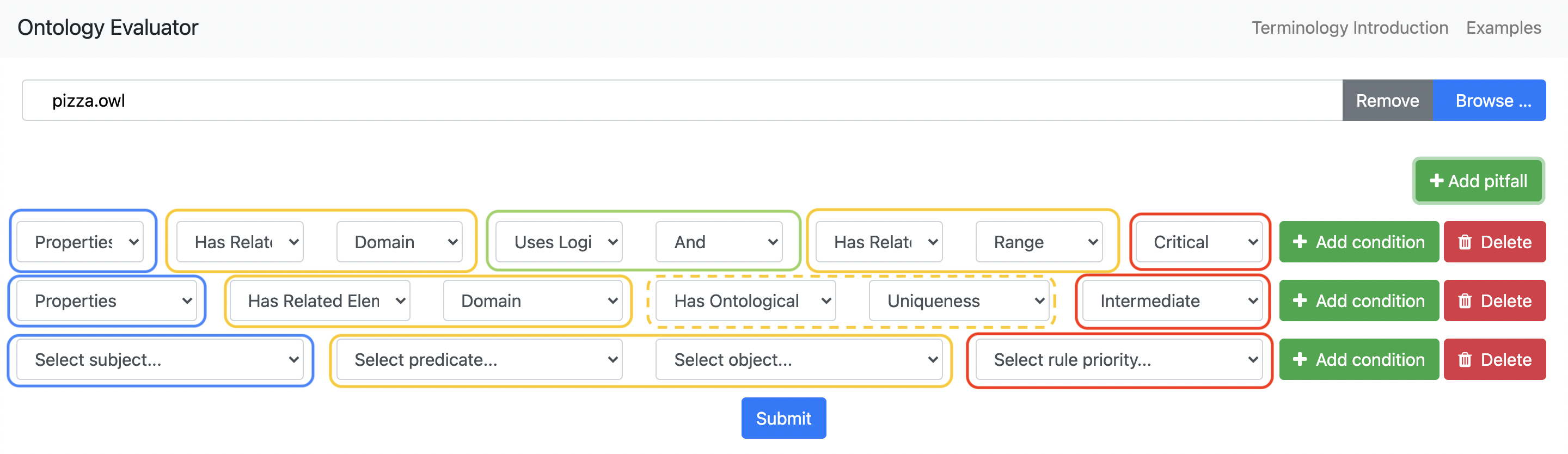}}
    \caption{The SynEvaluator interface. Subjects are highlighted in blue, Clauses in yellow, Operator Expressions in green and Rule Priorities in red. Among Clauses, Extractive Clauses are shown in solid lines while Functional Clauses are outlined with dotted lines.}
    \label{fig:interface}
\end{figure}

\begin{figure*}[h!]
    \centering
    {\includegraphics[width=\textwidth,keepaspectratio]{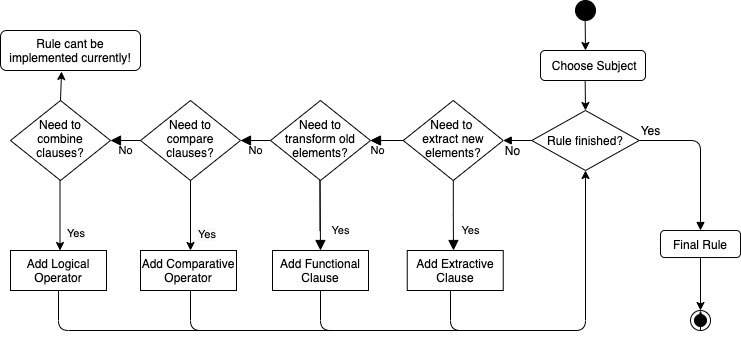}}
    \caption{Activity Diagram for creating a rule in SynEvaluator}
    \label{fig:flowchart-SynEvaluator}
\end{figure*}

The rule creation framework proposed above is used to create a web application for use by the ontology engineer. The primary interface of this application, SynEvaluator, is shown in Figure \ref{fig:interface}. It allows the users (ontology engineers) to use dropdown menus to create functional and extractive clauses, operator expressions and thus, rules, as well as set priorities for these rules. Users can choose between `Low', `Medium' and `High' priorities based on the task-specific importance of the rule. Figure \ref{fig:flowchart-SynEvaluator} shows, with the help of an activity diagram, how a user could create an appropriate rule using SynEvaluator. Finally, after uploading the enriched ontology and creating the rules, SynEvaluator parses the ontology using its parsing module and executes the created rules (Figure \ref{fig:workflowdiagram}). Post evaluation, the user is presented with a list of violating elements along with count and priority of each valid rule. 
\subsubsection{Benefits and Limitations}
The proposed framework makes it significantly easier for non-programmers to create customised rules dynamically. Also, compared to previous quality evaluation tools, due to the framework's ability to reuse keywords to create new rules, the developer effort required  to hard-code rules is minimised. Another major benefit is that the proposed framework can potentially be used to query over entire OWL language. This can be done as any ontological element/attribute can be extracted using extractive clauses and the appropriate function executed on them. Lastly, due to functional clauses, it is possible to execute ontological or ontology-level functions like normal query languages and use linguistic analysis on text. This is particularly useful in quality evaluation while applying appropriateness checks on IDs or labels.

\begin{figure}[h!]
    \centering
    {\includegraphics[width=0.7\textwidth]{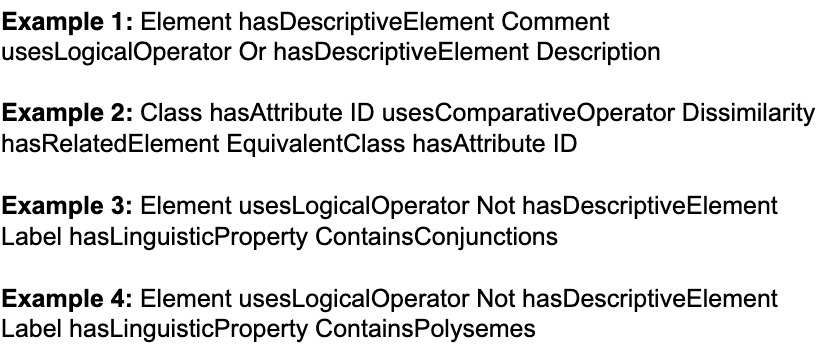}}
    \caption{More examples of rules supported by SynEvaluator}
    \label{fig:more-examples}
\end{figure}

In Figure \ref{fig:examples-rules} (Example 1), it can be seen that properties with missing domain or range, a common quality violation, could be detected through a combination of logical operators and extractive clauses. Another common violation is related to the absence of annotations (comments, descriptions, etc.) for an ontological element \cite{vrandevcic2009ontology}. This can be converted into a rule, once again through Extractive Clauses as shown in Figure \ref{fig:more-examples} (Example 1). A third quality violation would be when similar (or synonymous) classes are incorrectly defined as equivalent classes \cite{noy2001ontology}. This can be defined as in Figure \ref{fig:more-examples} (Example 2), where the comparative operator `Dissimilarity' is used to test for semantic similarity through cosine similarity of label embeddings. As mentioned previously, it is also possible to perform linguistic tests unlike other query languages. A basic example of this is detection of conjunctions in a label as shown in  Figure \ref{fig:more-examples}(Example 3). This is useful to identify quality violations where different concepts are merged in the same class using conjunctions \cite{poveda2012validating}. A more advanced example of a linguistic test would be a rule detecting polysemous elements (Figure \ref{fig:more-examples}, Example 4). This is useful in detecting violating elements that have labels denoting differing conceptual ideas in differing senses. SynEvaluator implements this check through the use of WordNet's synsets to find out how many senses a word can have. Once these rules are created, the ontology engineer can add domain/range elements; annotations; remove synonymous equivalent classes and fix classes with conjunctions and polysemes as appropriate. SynEvaluator can thus help in fixing structural and linguistic quality violations.

The current version of SynEvaluator has a few limitations.  It is currently only possible to chain clauses together or use operators to compare chained clauses. As a result, it is not possible to create rules with multiple lines. One major consequence of this is that variable assignment is not supported, and it is not possible to create a variable in one line and refer to it in another, as part of the same rule. Aggregation operators, such as `Count` or `Sum`, are currently not supported either. Finally, it is not possible to create rules that require reasoning. The described limitations shall be addressed in future iterations of SynEvaluator. In spite of the limitations, the current framework (as shown in Experiments section) is still able to support creation of the majority of quality evaluation rules.

\subsection{Stage 2: SemValidator}
SemValidator uses a crowdsourced approach to semantically evaluate ontologies. The key feature of SemValidator is that it does not require ontology engineers, domain experts or knowledge of OWL language for validation. This is useful in a crowdsourced setting, where validators may have varying degrees of expertise and knowledge. If further necessitates the use of appropriate quality control mechanisms. To ensure quality, SemValidator uses TweetExpert algorithm to calculate expertise score of a crowdsourced validator, which is then used to weigh their decisions. This section describes the approach used by TweetExpert and justifies the choices made. This is then followed by a discussion on the assumptions made by TweetExpert and the feasibility of the assumptions in the context of crowdsourcing. The section finally ends with a description of the implementation of SemValidator and how it can be used by crowdsourced validators.
%  \begin{figure}[h!]
%     \centering
%     {\includegraphics[width=\textwidth]{ontovalidator.png}}
%     \caption{The main validation interface of SemValidator. The Stanford Pizza ontology has been uploaded on SemValidator, with its concepts in blue and the enriched concepts in green. One of the enriched concepts, ``Tandoori Pizza" has been selected, with options to ``Accept" and ``Reject" it.}
%     \label{fig:SemValidator}
% \end{figure}
% As shown in Figure \ref{fig:processdiagram}, the ontology engineer responsible for evaluating the ontology through SynEvaluator (or more generally, an administrator) uploads it for semantic validation to SemValidator. Then, crowdsourced validators login to the application using their Twitter profiles and can then validate the enriched concepts, relations and instances of this ontology using a WebVOWL-based ontology visualization interface. Then accept/reject decisions are made for the enriched elements of the ontology, be it concepts, relations or instances, and these decisions are logged to a database. Once all validators have input their decisions, we calculate the expertise score of each validator using the TweetExpert algorithm proposed in Algorithm \ref{alg:tweetexpertise}. 
% We also illustrate the approach used by TweetExpert diagramatically in Figure \ref{fig:TweetExpert-algo}.

%  \begin{figure}[h!]
%     \centering
%     {\includegraphics[width=\textwidth]{TweetExpert-algorithm.jpeg}}
%     \caption{A process diagram illustrating the working of TweetExpert}
%     \label{fig:TweetExpert-algo}
% \end{figure}

\subsubsection{TweetExpert algorithm}
TweetExpert algorithm takes Twitter usernames of validators as input. For each of the validators, it calculates two scores: a) a `TweetSim' score and b) a `FriendSim' score. A `TweetSim' score is intended to assess the similarity of the validator's tweets to the domain of the ontology while `FriendSim' estimates the domain similarity of the validator's friends  (pages they are following). To calculate the `TweetSim' score, the validator's `n' most recent tweets are extracted from their profile and their semantic similarity with the domain keyword is computed using the Universal Sentence Encoder (USE) \cite{cer2018universal}. Here the `domain keyword' is manually chosen as the word most relevant to the domain of the ontology. For instance, ``Pizza" for Stanford Pizza ontology \cite{StanfordPizza} and ``Information Security" for the ISO 27001 based ``Information Security" ontology \cite{SteFenz}. The current implementation uses only one keyword, but it is possible to compute similarity scores with multiple keywords and average these scores for greater accuracy. The similarities are sorted in decreasing order. Out of n similarities, the top-K similarities are chosen. These top-K similarities are then averaged to yield `TweetSim' score. A similar approach is used to calculate `FriendSim' score. The `m' most recent friends are extracted to calculate their `TweetSim' scores. After sorting in decreasing order, the top-K most similar scores are averaged to yield `FriendSim' score. The reason behind extracting the most recent tweets and friends is to get a better estimate of the validator's current knowledge and interests. On the other hand, a top-K average helps in both filtering out occasional out-of-domain tweets from domain experts and smoothening out the effects of coincidental in-domain outliers from non-experts. The value of K is thus appropriately empirically chosen such that it is large enough to not include in-domain outliers from laymen, but small enough to exclude any out-of-domain tweets from experts. The pseudocode for TweetExpert is shown in Algorithm \ref{alg:tweetexpertise}.

\begin{algorithm*}[h!]
\SetAlgoLined
\KwResult{TweetExpert score}\
%  domain\_name, username\;
%  user\_tweets := extract\_tweets(username, max\_tweets)\;
%  tweet\_similarities := USE\_similarity(user\_tweets, domain\_name)\;
%  tweet\_similarities := decreasing\_sort( tweet\_similarities)[:max\_relevant\_tweets]\;
%  tweet\_score = average(tweet\_similarities)\;\
 
%  user\_friends := extract\_friends(username, max\_friends)\;
%  user\_friends\_tweets := extract\_tweets(user\_friends, max\_tweets)\;
%  friend\_tweet\_similarities := USE\_similarity(user\_friends\_tweets, domain\_name)\;
%  friend\_tweet\_similarities := decreasing\_sort(friend\_tweet\_similarities)[:max\_relevant\_tweets]\;
%  friends\_score = average(friend\_tweet\_similarities)\;\
%  score := tweet\_score, friends\_score\;
\DontPrintSemicolon
  \SetKwFunction{FMain}{calculate\_tweet\_similarity}
  \SetKwProg{Fn}{Function\,}{:}{}
  \Fn{\FMain{$username$}}{
        user\_tweets := extract\_tweets(username)\;
         tweet\_similarities := calculate\_USE\_similarity(user\_tweets, domain\_name)\;
        %  tweet\_similarities := sort\_decreasing(tweet\_similarities)\;
         best\_tweet\_similarities := get\_top\_K\_tweets(tweet\_similarities)\;
         tweet\_similarity\_score = average(best\_tweet\_similarities)\;
        \KwRet tweet\_similarity\_score\;
  }\;
  
  \SetKwFunction{FMain}{calculate\_friend\_similarity}
  \Fn{\FMain{$username$}}{
        user\_friends := extract\_friends(username)\;
        user\_friends\_scores := \lForEach{ $friend \in user\_friends$}{ calculate\_tweet\_similarity(friend)}
        best\_friend\_scores := get\_top\_K'\_friends(user\_friends\_scores)\;
        friend\_similarity\_score = average(best\_friend\_scores)\;
        \KwRet friend\_similarity\_score\;
  }\;
 
 tweet\_sim := calculate\_tweet\_relevance(username)\;
 friend\_sim := calculate\_friend\_relevance(username)\;
 score := ML\_predict\_score(tweet\_sim, friend\_sim)\;
 
%  friend\_tweet\_similarities := calculate\_USE\_similarity(user\_friends\_tweets, domain\_name)\;
%  friend\_tweet\_similarities := sort\_decreasing(friend\_tweet\_similarities)\;
%  best\_friend\_similarities := get\_top\_K\_tweets(tweet\_similarities)\;
%  friend\_relevance\_score = average(friend\_tweet\_similarities)\;\
 
%  score := tweet\_score, friends\_score\;

 \caption{TweetExpert algorithm}
 \label{alg:tweetexpertise}
\end{algorithm*}

Finally, the calculated `TweetSim' and `FriendSim' scores are input to a pre-trained Machine Learning regression algorithm that predicts the final $TweetExpert$ score using the two scores as feature vectors. The current implementation uses Epsilon-Support Vector Regression (SVR), since it was experimentally found to yield best results (shown in Experiments section). However, the system uses strategy software design pattern \cite{gamma1995design} which enables easy interchange of regression algorithms. The TweetExpert score is calculated for each of the validators by repeating the process described above. The final decision is taken using a weighted majority voting algorithm with the TweetExpert scores being used as weights. The TweetExpert scores provide a way to estimate a confidence value for the decisions input by each validator. This is particularly crucial as a quality control metric in non-probabilistic sampling techniques like crowdsourcing where number of validators can grow exponentially in count and diversity.
\subsubsection{Assumptions}
SemValidator makes reasonably grounded assumptions to establish efficacy and suitability in a crowdsourced setting. For example, about 49\% of crowdsourced workers whose primary source of income is Amazon Mechanical Turk, a popular crowdsourcing platform, are in the 18-29 age range\footnote{https://pewrsr.ch/3vX5q2G}. As of February 2021, about 42\% of Americans in the 18-29 age range use Twitter, with this age group being the most active demographic on Twitter\footnote{https://bit.ly/3z5IDn9}. It is also assumed that in order for expertise estimation to work, Twitter users tweet reasonably frequently. The average number of tweets per day per user, according to a 2016 study\footnote{https://bit.ly/3fSjWmA}, is 4.422, which translates to over 1600 tweets a year. This is typically sufficient since $K\sim50-100$. Finally, SemValidator assumes workers have public Twitter profiles, which would enable tweets and friends of a worker to be extracted. This assumption is predicated upon a recent 2019 survey\footnote{https://bit.ly/3pmMc3O} that showed 87\% of Twitter users in USA have public accounts. Also, as part of future work, SemValidator would allow login through Facebook and Linkedin as well and a similar algorithm for expertise estimation could be used. This is expected to further increase the applicability of this expertise-based approach for crowdsourcing.
\subsubsection{Proposing SemValidator: the final validation workbench}

 \begin{figure}[h!]
    \centering
    {\includegraphics[width=\textwidth]{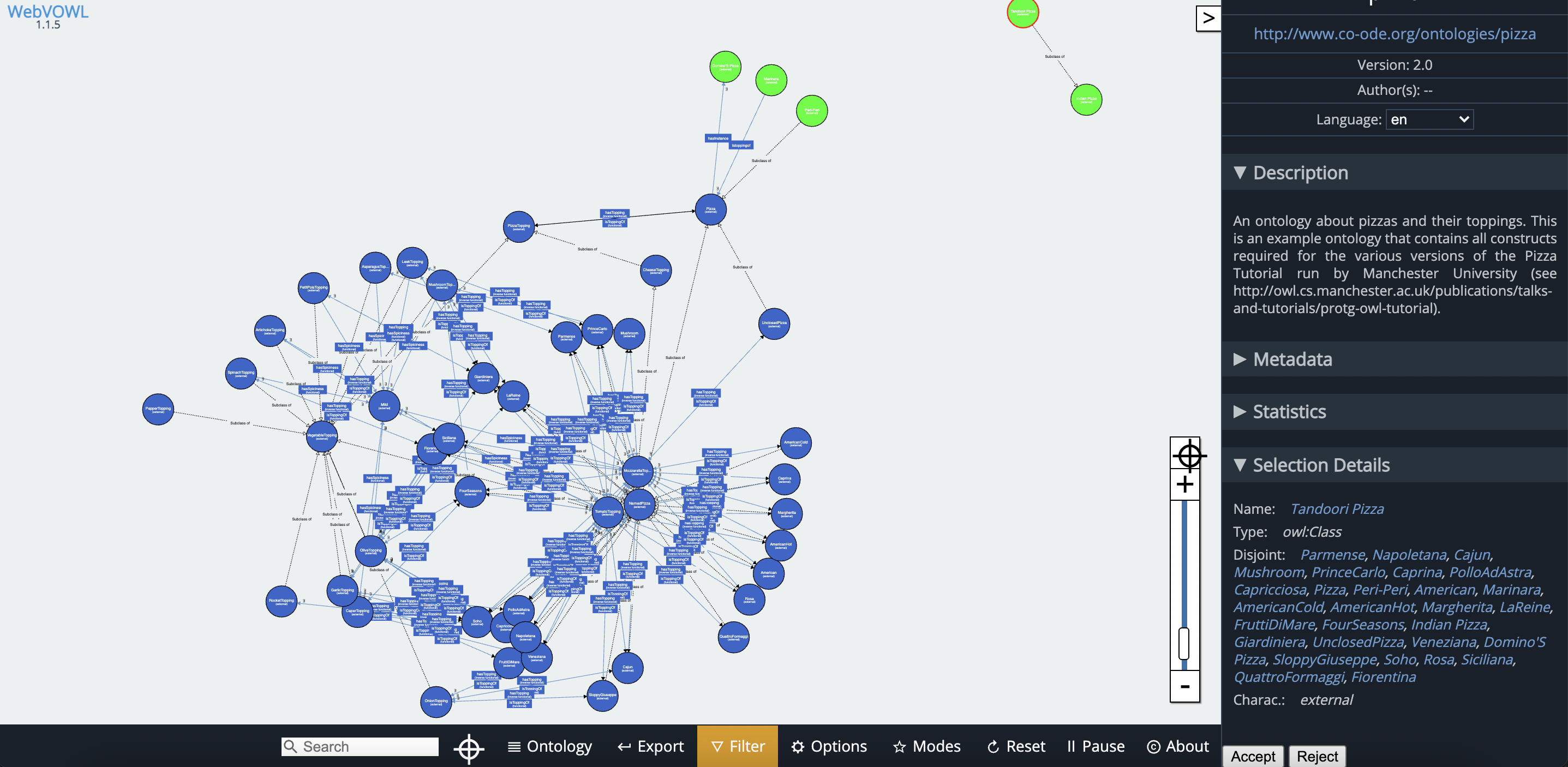}}
    \caption{The main validation interface of SemValidator. The Stanford Pizza ontology has been uploaded on SemValidator, with its concepts in blue and the enriched concepts in green. One of the enriched concepts, ``Tandoori Pizza" has been selected, with options to ``Accept" and ``Reject" it.}
    \label{fig:SemValidator}
\end{figure}

The proposed workflow is used to develop a validation workbench for crowdsourced workers that allows for accepting or rejecting enriched concepts, relations and instances in an enriched ontology. The main validation interface of this workbench, called SemValidator, is shown in Figure \ref{fig:SemValidator}. This application integrates Twitter authentication and uses the TweetExpert algorithm for calculating validator expertise. SemValidator allows for two types of users: (a) the administrator and (b) the validator. The administrator, typically the ontology engineer, can upload/delete ontologies to be validated and also access decisions made by validators. When the validator selects an ontology to validate, the ontology is served using the WebVOWL ontology visualization software. The enriched concepts and instances are highlighted in green, and on selection, enable the validators to select accept/reject decisions accordingly. Enriched relations, also in green, may also be accepted/rejected independent of the concepts they relate. The validators' decisions are recorded by SemValidator and logged to a database. The administrator can download this database after the crowdsourcing survey is completed, evaluate validator expertise using their Twitter usernames and then make final accept/reject decisions.

The syntactic and semantic evaluation aspects of the framework may be used independently of each other. However, utilizing the framework in an integrated manner as shown in Figure \ref{fig:workflowdiagram} is expected to give best results. This way, syntactical violations can be detected easily and in an automated and customizable manner, while semantic violations can now be detected more accurately by crowdsourced validators. The resulting ontology has enhanced syntactic and semantic ontological quality and is now fit for reuse.

%=============================================================
\section{Experiments}
Stanford Pizza and ISO-IEC 27001 Information Security ontologies are evaluated using the proposed framework to demonstrate ontology quality evaluation. Since the focus of this work is on enriched ontologies, these are manually enriched with concepts, properties and instances before quality evaluation. The RDF triples used to enrich Pizza and Information Security ontologies are provided over here\footnote{https://bit.ly/3z5lGR1}. Only 5 triples were chosen for this iteration, considering that ontology enrichment is as an iterative process and the count of triples per iteration is expected to be of this order. For Pizza, the domain-specific webpages used for extraction consisted of culinary articles\footnote{https://bit.ly/3yYGGZP} and food travel blogs\footnote{https://bit.ly/3iiEtCm}. For Information Security they consisted of informative articles and product pages by Cisco\footnote{https://bit.ly/3wXRHZj}, Barracuda\footnote{https://bit.ly/34PIUN9} and SearchSecurity\footnote{https://bit.ly/3wXRJQV}. Please note that some relevant data in this section is shown through external links due to space constraints.

% \begin{table}[]
% \centering
% \caption{RDF triples used to enrich pizza ontology along with their labels}
% \label{tab:pizza-dataset}
% \begin{tabular}{|c|c|c|c|}
% \hline
% Term A            & Term B        & Relation & Label \\ \hline
% Anchovies         & Pizza Topping & Subclass & True  \\ \hline
% Neapolitan        & Pizza         & Subclass & True  \\ \hline
% Basil             & Pizza Topping & Subclass & True  \\ \hline
% Quad Cities Pizza & Italy & Originates from & False \\ \hline
% Greek Pizza       & Pizza         & Subclass & True  \\ \hline
% \end{tabular}

% \end{table}

% \begin{table}[]
% \centering
% \caption{RDF triples used to enrich security ontology along with their labels}
% \label{tab:security-dataset}
% \begin{tabular}{|c|c|c|c|}
% \hline
% Term A             & Term B           & Relation & Label \\ \hline
% Botnet             & Malware          & Subclass & False \\ \hline
% Firewall           & Network Security & Subclass & True  \\ \hline
% PII                & Data             & Subclass & True  \\ \hline
% Barracuda CloudGen & Firewall         & Instance & True  \\ \hline
% ILoveYou           & Ransomware       & Instance & False \\ \hline
% \end{tabular}
% \end{table}

\subsection{Syntactic Quality Evaluation using SynEvaluator}

% In this section, the results of Ontology evaluation using SynEvaluator are detailed. The major difference between SynEvaluator and other currently available tools is its wide applicability, flexibility and violation-detection accuracy. 
This section attempts to evaluate the applicability and accuracy of SynEvaluator in creating rules and detecting violations respectively.
It is hypothesized that SynEvaluator's reusable theoretical framework for priority-specific rule creation increases its applicability for diverse range of tasks, but without compromising on violation-detection accuracy.
% Additionally, the accuracy of detecting violations in ontologies based on the implemented rules is a major differentiating criteria. 
To prove these claims, SynEvaluator is compared against OOPS! \cite{poveda2018oops}, a popular, publicly available SOTA tool that allows qualitative rule evaluation using rules. Moreover, OOPS! compiles 41 most commonly observed pitfalls drawn from several popular works on ontology quality evaluation \cite{gomez1999evaluation}, \cite{noy2001ontology}, \cite{vrandevcic2009ontology} and the ones described in the OOPS! catalogue \cite{oopscatalogue} are chosen as rules for implementation. This allows appropriate assessment of applicability for rule creation, while also allowing accuracy evaluation by comparing SynEvaluator's detected violations to that of OOPS! Note that while SynEvaluator's suitability for rule creation is being assessed here by comparing with rule-based evaluation approaches, it cannot be compared yet with other contemporary works focusing on metric-based evaluation. This is being planned as part of future work by introducing metric-to-rule conversion, which would allow metrics to be framed as rules. One possible way this could be done is by adding conditions, such as comparison operators, after metrics to form rules.
% Extending SynEvaluator to support quantitative metrics and comparing with other works is being planned in future. 

\begin{table}[]
\centering
\caption{No of pitfalls implemented by SynEvaluator (SE) Vs OOPS!}
\setlength\tabcolsep{3pt}
\begin{tabular}{|c|c|c|}

\hline
                                                                  \diagbox{SE}{OOPS!}
 & \begin{tabular}[c]{@{}c@{}}Implemented\end{tabular} & \begin{tabular}[c]{@{}c@{}}Not implemented \end{tabular} \\ \hline
\begin{tabular}[c]{@{}c@{}}Implemented\end{tabular}     & 29                                                                     & 4                                                                         \\ \hline
\begin{tabular}[c]{@{}c@{}}Not implemented \end{tabular} & 1                                                                      & 7                                                                         \\ \hline
\end{tabular}
\label{tab:pitfall-comparison}
\end{table}

SynEvaluator's framework can be used to successfully formulate 30 of the 41 pitfalls compiled in the catalogue, compared to OOPS! which can automate 33. As shown in Table \ref{tab:pitfall-comparison}, the majority of pitfalls (29 of 41) can be successfully implemented by both OOPS! and SynEvaluator. 7 cannot be implemented by either of them, and this involves rules that require human knowledge and reasoning (such as overspecialization of ontology hierarchy, usage of wrong relations, etc.). 1 of the pitfalls (involving linguistic polysemy detection) could be implemented by SynEvaluator but not by OOPS!. There are 4 pitfalls implemented by OOPS! which SynEvaluator cannot check. This includes rules that check for undeclared disjoint concepts, undeclared transitive properties, equivalent classes, etc. Supporting such rules involves a higher degree of ontological reasoning that SynEvaluator is incapable of. Nevertheless, SynEvaluator's ability to implement vast majority of quality evaluation rules off-the-shelf in a customizable manner increases confidence in its applicability for future evaluation tasks that may involve more rules.
% In addition, due to the ability to reuse keywords, SynEvaluator facilitates creation of a far greater number of rules than could be achieved initially through hard-coding of individual rules. This in turn, ends up significantly minimising developer effort in programming rules as well. 

For comparing accuracy of violation detection, SynEvaluator's reported violations are compared with OOPS! \cite{poveda2018oops}. Given the absence of ground truth and the widespread popularity of OOPS!, it is assumed that OOPS! uses valid, non-erroneous rule checks for the implemented pitfalls. It can be observed that the violations reported by SynEvaluator match those of OOPS! in all pitfalls except for P22. OOPS! mentions that these could be due to inconsistent naming conventions, however SynEvaluator is unable to detect any such errors in the ontologies. This may be a result of the incorrect naming checks currently used.  But for the other pitfalls, both tools returning the same number of violations suggests that for the implemented framework, SynEvaluator is accurate in evaluating rules and detecting pitfalls.
 
% Thus, for the pitfalls where both tools, despite using two different methodologies report the same violations for both pizza and security ontologies, we can reasonably assume the accuracy of the checks performed by SynEvaluator. 
% SynEvaluator is able to automatically evaluate ontologies for P1, though OOPS! is unable to do the same. Having to  SynEvaluator is able to factor nearly all the rules automatically checked by OOPS!, except pitfalls P24, P29 and P30 as rules. 
% For the rest, namely the pitfall checks implemented by both OOPS! and SynEvaluator, both return the same number of violations. This, in turn, proves the accuracy of SynEvaluator in evaluating rules and detecting pitfalls.

% Our experiments show that our approach, particularly our theoretical framework, seems viable for implementation of popular quality evaluation rules, also that the detected violations are valid. 

% \usepackage{multirow}

\subsection{Semantic Quality Validation using SemValidator}

SemValidator uses a crowdsourced survey to semantically validate the  enriched Pizza and Information Security ontologies. Crowdsourced validators were invited by tweeting a description of the task along with the link to the SemValidator application. Among the 31 validators that participated in the survey, 3 had private Twitter accounts so their responses were discarded. The 28 validators with public accounts included 5 validators with a background in Cybersecurity, 4 in Artificial Intelligence, 5 in Software Engineering, 4 in Healthcare, 3 in Food domain (such as chefs and culinary experts), 3 in Literature, 2 in Politics and 1 each in Fashion and Pure Science. These diverse bunch of validators were asked to self-identify their domain of expertise for statistical purposes before participating in the survey.  To preserve validator anonymity, the anonymized results of the survey are provided  here\footnote{https://bit.ly/3pvzbFh}. After completion of the survey for both ontologies, `TweetSim' and `FriendSim' scores were calculated for all 28 validators using the TweetExpert algorithm, described previously. TweetExpert took \~15 seconds to execute for our values of $K=20$ and $K'=5$ per validator profile. Finally, TweetExpert score was calculated by experimenting with 4 standard regression algorithms: Linear Regression, Random Forest Regression, K-Nearest Neighbours (KNN) Regression and Epsilon-Support Vector Regression (SVR). A standard majority voting where all decisions have equal weightage is considered as a naive baseline. 

The input features for each algorithm were `TweetSim' and `FriendSim' scores of a validator and the label was the ratio of correct answers (to total answers) given by them. This was done to ensure validators with more correct answers received higher `TweetExpert' scores using similarity scores as feature vectors. Since a gold standard for the correct answers did not exist, CISOs were asked to manually validate the enriched triples from the information security domain, while the authors themselves validated the triples from the pizza domain. To determine the best among the 4 regression algorithms for score prediction, 7-fold cross validation was carried out. The 28 responses (now reduced to feature vectors and labels) available for both ontologies were divided into 7 folds of 4 responses each. 5 folds were used for training, 1 for validation and 1 for testing. For each chosen test fold, immediately succeeding fold was chosen as the validation set and all the other folds were used for training. 

% \begin{table}[]
% \centering
% \caption{Hyperparameters experimented for each algorithm}
% \label{tab:hyperparams}
% \begin{tabular}{|c|c|}
% \hline
% Algorithm         & Hyperparameters                                                            \\ \hline
% Majority Voting & N/A 
%                     \\ \hline
% Linear Regression & N/A                                                                          \\ \hline
% Random Forest     & n\_estimators = {[}10, 50, 100{]}                                          \\ \hline
% KNN               & n\_neighbors = {[}3,5,7,9{]}, weights = {[}"uniform", "distance"{]}        \\ \hline
% SVR               & C = {[}1, 10, 50, 100, 500{]}, epsilon = {[}0.001, 0.05, 0.01, 0.1, 0.5{]} \\ \hline
% \end{tabular}
% \end{table}

Different sets of hyperparameters were considered for each regression algorithm\footnote{https://bit.ly/3wUa9SA}. The best set of hyperparameters for each algorithm was determined by selecting the algorithm with highest accuracy on the validation set, calculated by determining the proportion of correctly answered questions to the total number of questions. The predicted answer to a question is determined by using weighted majority voting algorithm on the predicted expertise scores in the validation set. The higher the number of correct answers, the closer the predicted scores are to their actual value. The same procedure was followed for calculating the accuracy on the test folds. Accuracy obtained on test and validation sets for each of these algorithms is shown in Table \ref{tab:results-regression} for both ontologies. Epsilon Support Vector Regression was found to yield best results on both datasets.

To determine whether training score prediction algorithms requires feature vectors from the same domain, an SVR model is trained on a more generic dataset, i.e the Pizza dataset and tested on Information Security, a niche dataset. The model hyperparameters are chosen as the ones that performed best on the Information Security dataset earlier. Varying percentages of the Pizza dataset are used to construct training subsets, to observe variation in test accuracy with subset size. To ensure randomness while choosing these subsets, 10 experimental trials were conducted where the dataset is shuffled before each trial and trained on the first $x\%$ of samples, where $x$ is the percentage of the training dataset chosen. The results are shown in Table \ref{tab:percentage-accuracy}. The accuracy monotonically increases with increase in size of training subset. The results suggest that a model pre-trained to predict the expertise score on one domain could be reused multiple times in different domains, even if they are niche.

% \ usepackage{multirow}
\begin{table}[]
\centering
\caption{Mean Validation and Test Accuracy scores on Pizza and Security datasets}
\label{tab:results-regression}
\begin{tabular}{|c|c|c|c|c|}
\hline
\multirow{2}{*}{Algorithm} & \multicolumn{2}{c|}{Pizza} & \multicolumn{2}{c|}{Security} \\ \cline{2-5} 
 & Val & Test & Val & Test \\ \hline
Majority Voting & 71.43 & 71.43 & 51.43 & 51.43 \\ \hline
Linear Regression & 85.71 & 83.57 & 68.57 & 71.43 \\ \hline
Random Forest & 80.0 & 81.43 & 68.57 & 74.28 \\ \hline
KNN & 85.71 & 80.0 & 77.14 & 77.14 \\ \hline
SVR & \textbf{88.57} & \textbf{85.71} & \textbf{80.0} & \textbf{80.0} \\ \hline
\end{tabular}
\end{table}

%==================================
% \begin{table}[h!]
% \caption{Results on running 4 ML regression algorithms on the security dataset}
% \label{tab:security-results}
% \begin{tabular}{|c|c|c|c|}
% \hline
% Algorithm & \begin{tabular}[c]{@{}c@{}}Optimal\\ Hyperparameters\end{tabular} & Accuracy (val) & Accuracy(test) \\ \hline
% Majority Voting & N/A & 51.43 & 51.43 \\ \hline
% Linear Regression & N/A & 68.57 & 71.43 \\ \hline
% Random Forest & n\_estimators = 10 & 68.57 & 74.28 \\ \hline
% KNN & n\_neighbors = 7, weights = 'uniform' & 77.14 & 77.14 \\ \hline
% SVR & C = 10, epsilon = 0.001 & \textbf{80.0} & \textbf{80.0} \\ \hline
% \end{tabular}
% \end{table}
%===========================
\begin{table}[]
\centering
\caption{Mean accuracy scores on test dataset (Information Security) with variation in \% of training dataset (Pizza).}
\label{tab:percentage-accuracy}
\begin{tabular}{|c|c|c|c|c|c|}
\hline
Dataset\% & 10\% & 25\% & 50\% & 75\% & 100\% \\ \hline
Accuracy      & 70\% & 76\% & 86\% & 92\% & 100\% \\ \hline
\end{tabular}
\end{table}

Table \ref{tab:results-regression} shows that the best-performing algorithm (TweetExpert with SVR followed by weighted majority voting) significantly outperforms naive majority voting. It can be observed that naive majority voting gives particularly bad results for Information Security, a relatively niche domain, than Pizza, a domain known more to laymen as well. When replacing naive majority voting with TweetExpert + SVR, a drastic increase in performance in Information Security (55.55\%) vs Pizza (20\%) can be observed. Even the worst performing regression algorithm gives an increase of 38.8\% and 14\% respectively. As a result, it can be inferred that estimating expertise using TweetExpert can be particularly useful for quality control in niche domains.
% The improvement is 55.55\% for information security and 20\% for the pizza domain: in fact, even TweetExpert with the worst performing regression algorithm outperforms majority voting by 14\% and 38.8\% respectively. 

% It needs to be noted that even the worst performing ML algorithm was found to show a substantial test accuracy increase for Information Security dataset when compared with naive majority voting.  This also means domains like cybersecurity are more prone to poor quality results when using crowdsourced validators, which in turn necessitates expertise estimation. This is further validated when the accuracies are compared using TweetExpert + SVR versus naive majority voting. A much higher increase in performance in the information security dataset (55.55\%) versus the pizza dataset (20\%).  Moreover, our experiment involving training on varying percentages of pizza dataset show that mean accuracy steadily increases with increase in size of training dataset, increasing to 100\% when all 28 tweets are used for training. This is even more encouraging when one considers the mentioned nicheness of the test dataset (cybersecurity), where it is significantly harder to get good quality responses.

\section{Conclusion and Future Work}
Ontology evaluation is a critical stage of any ontology engineering process. In this paper, SynEvaluator and SemValidator are proposed for syntactic and semantic evaluation of an enriched ontology respectively. Although the work focuses on an enriched ontology, it can be easily extended for evaluation of any generic ontology. 

SynEvaluator automatically evaluates ontologies using customised rules created dynamically at runtime, and improves on previous quality evaluation approaches in a variety of ways. Firstly, it offers greater flexibility to the user in terms of creating, updating and deleting custom rules as well as setting priorities. Secondly, it proposes rule creation by the usage of a novel theoretical framework that factors rules into sequences of `clauses' and `operator expressions'. This facilitates creation of an interactive interface that makes it easier for non-programmers to dynamically create rules. In addition, chaining together independent keywords can facilitate creation of a large number of rules without requiring additional developer programming.  SemValidator improves on previously proposed crowdsourced ontology validation approaches by incorporating a Twitter-based quality control mechanism. The TweetExpert algorithm is proposed for calculating the expertise score of a validator using the tweets and friends extracted from their Twitter profile as input. 

The efficacy of SynEvalautor was shown by implementing rules to detect pitfalls and the accuracy of the detected violations was compared with publicly available OOPS! tool. Semantic quality evaluation using SemValidator is performed on both Pizza and Information security ontologies with the help of a crowdsourced survey of 28 validators. The experimental results showed a significantly higher than naive majority voting. 

The experimental results are encouranging but also can be used as an aid to further extend the research work. For example, SynEvaluator can be expanded to support additional, more complex operations such as using parantheses, arithmetic operators, variable assignment etc. Although the TweetExpert algorithm used by SemValidator currently calculates expertise using only tweets and friends as feature vectors, it could be extended to use additional information such as Twitter lists, followers, tweet metadata, etc.  
\bibliographystyle{splncs04}
\bibliography{sample}

\end{document}